\documentclass[12pt]{iopart}

\usepackage{graphicx}
\usepackage{bm}
\usepackage{color}
\usepackage{rotating}
\begin{document}

\title{Coherent-Light Boosted, Sub Shot Noise, Quantum Interferometry}

\author{William~N.~Plick$^{1}$, Jonathan~P.~Dowling$^{1}$, and Girish~S.~Agarwal$^{2}$}

\address{$^{1}$Hearne Institute for Theoretical Physics, Department of Physics and Astronomy, Louisiana State University, Baton Rouge, LA 70803.\\
$^{2}$Department of Physics, Oklahoma State University, Stillwater, OK 74078}

\begin{abstract}
We present in this paper a new scheme for optical interferometry. We utilize coherent-beam-stimulated two-mode squeezed light, which interacts with a phase shifter and is then squeezed again before detection. Our theoretical device has the potential to reach far below the shot noise limit (SNL) in phase sensitivity. This new proposal avoids the pitfalls of other setups, such as difficulty in creating the required resource. Furthermore, our scheme requires no complicated detection protocol, relying instead only on simple intensity measurement. Also, bright, coherent sources ``boost" squeezed light, creating a very sensitive device. This hybrid scheme relies on no unknown components and can be constructed with current technology. In the following we present our analysis of this relatively straightforward device, using the operator propagation method. We derive the phase sensitivity and provide a simple numerical example of the power of our new proposal. Sensitivity scales as a shot noise limited Mach-Zehnder Interferometer, multiplied by a sub-Heisenberg contribution from the squeezed light.
\end{abstract}

\maketitle

\section{Introduction}
Accurate measurement is the cornerstone of our understanding of the natural world. Science has improved apace with our ability to examine ever more minute, remote, or subtle phenomena.

Optical interferometry is an extremely useful and flexible measuring tool. The Mach-Zehnder Interferometer (MZI) \cite{Z,M} and its cornucopia of variants have for many years been the standard devices used in this capacity. Optical interferometers function by splitting light between two paths, directing one path towards an object to be studied, recombining the beams, and observing the resultant interference pattern. This generic ``object" may be any number of things from a medium, which changes the properties of the light in the presence of magnetic fields \cite{Budker}, to a path difference caused by the presence of a gravitational distortion \cite{russ,weiss}. The power of optical interferometry comes partially from its very broad applicability. It also comes from its capacity for extreme sensitivity. 

The phenomenon we studied is represented by an abstract phase shifter placed in the detection arm of the interferometer. The measuring power of a given interferometric scheme can then be characterized by its sensitivity to changes in this phase.

With solely coherent light input into an interferometer the limit on the sensitivity to this abstract phase is $\Delta\phi^{2}=1/N_{\mathrm{Coh}}$. Where $N_{\mathrm{Coh}}$ is the average number of photons in the coherent beam. This root-intensity scaling is referred to as the shot noise-limit (SNL). In the past few decades it has been shown that this limit may be surpassed by taking advantage of the quantum nature of light. Viewing entanglement as a resource for sensitivity enhancement there has been much progress towards achieving the more fundamental Heisenberg limit which scales as $\Delta\phi^{2}=1/N^{2}$, where $N$ is (usually) the total or average number of particles in-putted into the interferometer. There are several main thrusts in this effort: Utilizing squeezed light as one or both inputs \cite{caves,kimble,os,luis,kolkiran1,martini,hoffloss} (or more recently placing the squeezer in the device itself \cite{NdM}), creating maximally path entangled number states (N00N states) inside the device \cite{boto,agarwal,zei,steinberg,nagata,ono,white,jon,gao,sean,ryan,vitelli,Silb}, use of Bose-Einstein condensates \cite{boixo2,est,boixo1}, causing the light to execute multiple passes through the phase shift \cite{higgins2,kitten,jeremy}, and other schemes \cite{holland,artur,qbs,qbs2,lee,kim}.

Though all these programs show promise, they all suffer from daunting problems in the technical implementation: difficulty in creating the required resource (i.e. creating high-N N00N states, large Bose-Einstein condensates, large-gain squeezed light sources, or coaxing the light into passing through the phase shift many times), exotic detection schemes (such as parity \cite{p1,p2,p3,p4,p5}), and reliance on yet to be developed technological elements.

As we shall show, our setup avoids many of these problems; whilst granting a large improvement in sensitivity. This flexibility comes from the fact that we allow for the interferometer to utilize bright coherent sources, which enhance the effect of squeezed light. In other words our scheme mitigates the difficulty in creating the super-sensitive resource by a allowing that resource to be amplified by readily available laser light. Furthermore our scheme uses only simple total intensity measurement, and could be built with current technology. 

We will take as an example the LIGO (Laser Interferometer Gravitational-Wave Observatory) project, and show that by using a squeezing parameter of $r=3$ (which is not far outside the realm of what is currently available) the shot noise may be reduced by a factor of two hundred. Alternatively the coherent intensity could be reduced by a factor of forty thousand, while maintaining the original sensitivity. Thus our new scheme has the potential to both improve the most sensitive devices and to make those devices more easily accessible.

\section{Interferometric Setup}

In 1985 Yurke, McCall, and Klauder \cite{klauder}, building on foundational work by W\'{o}dkiewicz and Eberly \cite{eb}, introduced a new class of interferometers which, unlike MZI and Fabry-Perot setups, is described by the group SU(1,1) \--- as opposed to SU(2). A realization of these devices may be imagined by taking a traditional MZI and replacing the 50-50 beam splitters with four-wave mixers. Klauder, et al., showed that the sensitivity of this device exhibits sub-Heisenberg scaling. Building on this work Sanders et al. completed a general analysis of these ``active interferometers'' using information theory, showing again the enhanced scaling \cite{Sanders}. Recently Kolkiran and Agarwal studied-coherent-beam stimulated parametric downconversion as an input into a MZI \cite{kolkiran2}. Their work demonstrated the ability to acquire resolution enhancement at high signal values with good visibility. 

In this work we modify the Klauder, et al., setup by replacing the vacuum inputs in the initial four-wave mixer with coherent states. In our treatment the four-wave mixers are expressed as optical parametric amplifiers. The \--- potentially very bright \--- coherent light in the interferometer ``boosts" the squeezed light from the optical parametric amplifiers (OPAs) into the high intensity regime, while maintaining sub-SNL scaling. For some references on injecting photons from an OPA into an external non-vacuum mode see for example Refs. \cite{in1,in2,in3}.

\begin{figure}\centering
\includegraphics[scale=0.3]{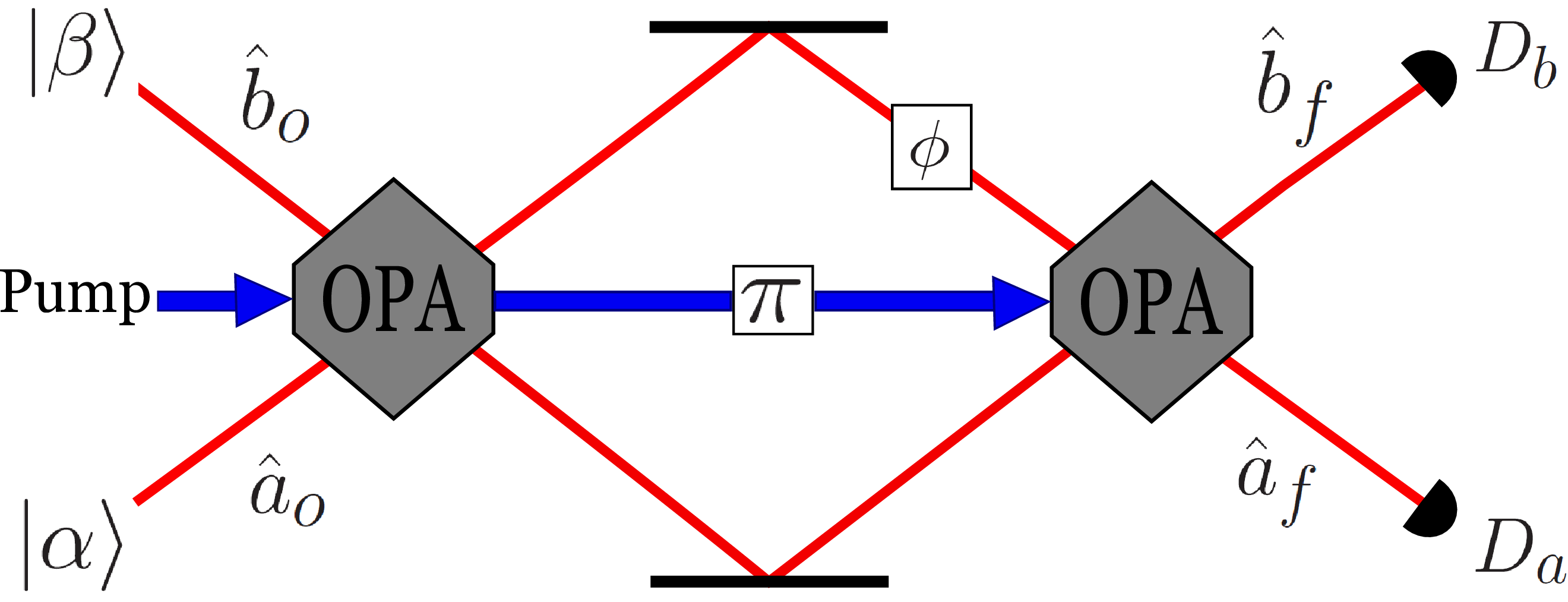}
\caption{\label{setup}A strong laser beam pumps the first OPA. The beam (which is assumed to be undepleted) undergoes a $\pi$ phase shift and then pumps the second OPA. The input modes of the first OPA are fed with coherent light. After the first OPA one of the outputs interacts with the phase to be probed. Both outputs are then brought back together as the inputs for the second OPA. Measurement is done on the second OPA's outputs.}
\end{figure}

Take a setup as depicted in Fig. \ref{setup}. An OPA is pumped by a coherent source. Assuming the pump beam is undepleted after the first OPA, it then undergoes a $\pi$-phase shift and pumps a second OPA. The first OPA is seeded with a coherent beam in each of its input modes. The phase shift to be probed is placed in the upper arm of the device and interacts with one of the output modes of the first OPA. Both output modes are then brought back together into the input modes of the second OPA. The output of the second OPA is then read out by a detector in each of the output modes. If the input ports $|\alpha\rangle$ and $|\beta\rangle$ are replaced by vacuum then the scheme reduces to that of Klauder et al.

This device can be modelled with the transformation between input modes and output modes given by $\vec{V}_{f}=\hat{M}_{3}\hat{M}_{2}\hat{M}_{1}\cdot\vec{V}_{o}$, where

\begin{eqnarray}
\vec{V}_{f}&\equiv &\left[\begin{array}{c}\hat{a}_{f}\\\hat{a}_{f}^{\dagger}\\\hat{b}_{f}\\\hat{b}_{f}^{\dagger}\end{array}\right], \quad \vec{V}_{o}\equiv\left[\begin{array}{c}\hat{a}_{o}\\\hat{a}_{o}^{\dagger}\\\hat{b}_{o}\\\hat{b}_{o}^{\dagger}\end{array}\right], \quad
\hat{M}_{1}\equiv\left[\begin{array}{cccc}\mu&0&0&\nu\\0&\mu&\nu&0\\0&\nu&\mu&0\\\nu&0&0&\mu\end{array}\right]\nonumber\\
\hat{M}_{2}&\equiv &\left[\begin{array}{cccc}e^{i\phi}&0&0&0\\0&e^{-i\phi}&0&0\\0&0&1&0\\0&0&0&1\end{array}\right],\quad
\hat{M}_{3}\equiv\left[\begin{array}{cccc}\mu&0&0&-\nu\\0&\mu&-\nu&0\\0&-\nu&\mu&0\\-\nu&0&0&\mu\end{array}\right],\nonumber\nonumber
\end{eqnarray} 

\noindent and $\mu=\cosh{r}$, $\nu=\sinh{r}$, $r$ is the gain of the the OPAs, and the phase of the initial pump has been set to zero. The matrix $\hat{M}_{1}$ represents the first OPA, $\hat{M}_{2}$ represents the phase shift, and $\hat{M}_{3}$ represents the second OPA. Note that $\hat{M}_{3}\hat{M}_{1}=\hat{I}$. Thus in the case that $\phi=2\pi n$, $n=0,1,2...$ the input equals the output. 

We wish to analyse the sensitivity of this setup to the phase $\phi$; given a simple total intensity detection scheme represented by the operator $\hat{N}_{T}=\hat{a}_{f}^{\dagger}\hat{a}_{f}+\hat{b}_{f}^{\dagger}\hat{b}_{f}$. We use the standard formula for phase sensitivity $\Delta\phi^{2}=\left(\langle\hat{N}_{T}^{2}\rangle-\langle\hat{N}_{T}\rangle^{2}\right)/\left(\partial_{\phi}\langle\hat{N}_{T}\rangle\right)^{2}.$ In order to perform this calculation we require the first and second moment of $\hat{N}_{T}$. The \textit{NCAlgebra} package \cite{NC} for $\mathrm{Mathematica}^{\mathrm{TM}}$ was utilized to create a program specially designed to do this. The result in our case is a series of rather large and un-illuminating equations, which are not reported here. However they may be used to calculate the phase sensitivity, given in by,

\begin{eqnarray}
\Delta\phi^{2}&=&\frac{1}{\Gamma}\mu^{2}\nu^{2}\left\lbrace B\left[1+4\cos(\phi)+3\cos{(2\phi)}+8\cosh(8r)\sin^{4}\left(\frac{\phi}{2}\right)\right.\right.\nonumber\\
& &\left.+8\cosh(4r)\sin^{2}(\phi)\right]+32|\alpha\beta|^{2}\left\lbrace\sin(2\theta)\sinh(2r)\left[2\cosh^{2}(2r)\sin(\phi)\right.\right.\nonumber\\ 
& &\left.-\sin(2\phi)\sinh^{2}(2r)\right]+2\cos(2\theta)\sin^{2}\left(\frac{\phi}{2}\right)\sinh(4r)\nonumber\\ & &\left.\times\left.\left[\cosh^{2}(2r)-\cos(\phi)\sinh^{2}(2r)\right]\right\rbrace-8\right\rbrace,\label{full}
\end{eqnarray}

\noindent where

\begin{eqnarray}
\Gamma&=&256\left\lbrace|\alpha\beta|\left[\mu^{2}\sin(2\theta+\phi)-\nu^{2}\sin(2\theta-\phi)\right]\right.\nonumber\\
& &+\left.\left[1+|\alpha|^{2}+|\beta|^{2}\right]\mu\nu\sin(\phi)\right\rbrace^{2}\nonumber
\end{eqnarray}

\noindent and $\theta$ is the phase of the input coherent states (which are taken to be equal), and $B=1+2|\alpha|^{2}+2|\beta|^{2}$. It should be noted that, while this calculation was carried out with computer aid, the results are analytical as the calculation was done symbolically \--- not numerically. We can check this formula by taking limits and comparing to known expressions. In the limit of $\alpha=\beta=\phi\rightarrow 0$ we obtain $\Delta\phi^{2}=1/\sinh^{2}(2r)$, which matches the result from Klauder, et al. A simplification of Eq. (\ref{full}) occurs when $\phi=0$ and $\theta=\pi/4$

\begin{eqnarray}
\Delta\phi^{2}=\frac{1}{N_{\mathrm{OPA}}(N_{\mathrm{OPA}}+2)}\frac{1}{N_{\mathrm{Coh}}},\label{simple}
\end{eqnarray} 

\noindent where the intensity $N_{\mathrm{Coh}}$ is the amount of coherent light at the input ($|\alpha|^{2}+|\beta|^{2}$), and the intensity $N_{\mathrm{OPA}}=2\sinh^{2}(r)$ is the amount of light the OPA would emit with vacuum inputs. We have taken $|\alpha|=|\beta|$ for the sake of simplicity. This setup multiplies the sub-Heisenberg sensitivity of the Klauder setup with the standard-quantum-limited sensitivity of a coherent light input MZI. The advantage provided by the coherent-light boosting is evident, allowing sub-SNL scaling at intensities far beyond what entangled sources alone can provide.   

\begin{figure}\centering
\includegraphics[scale=0.37]{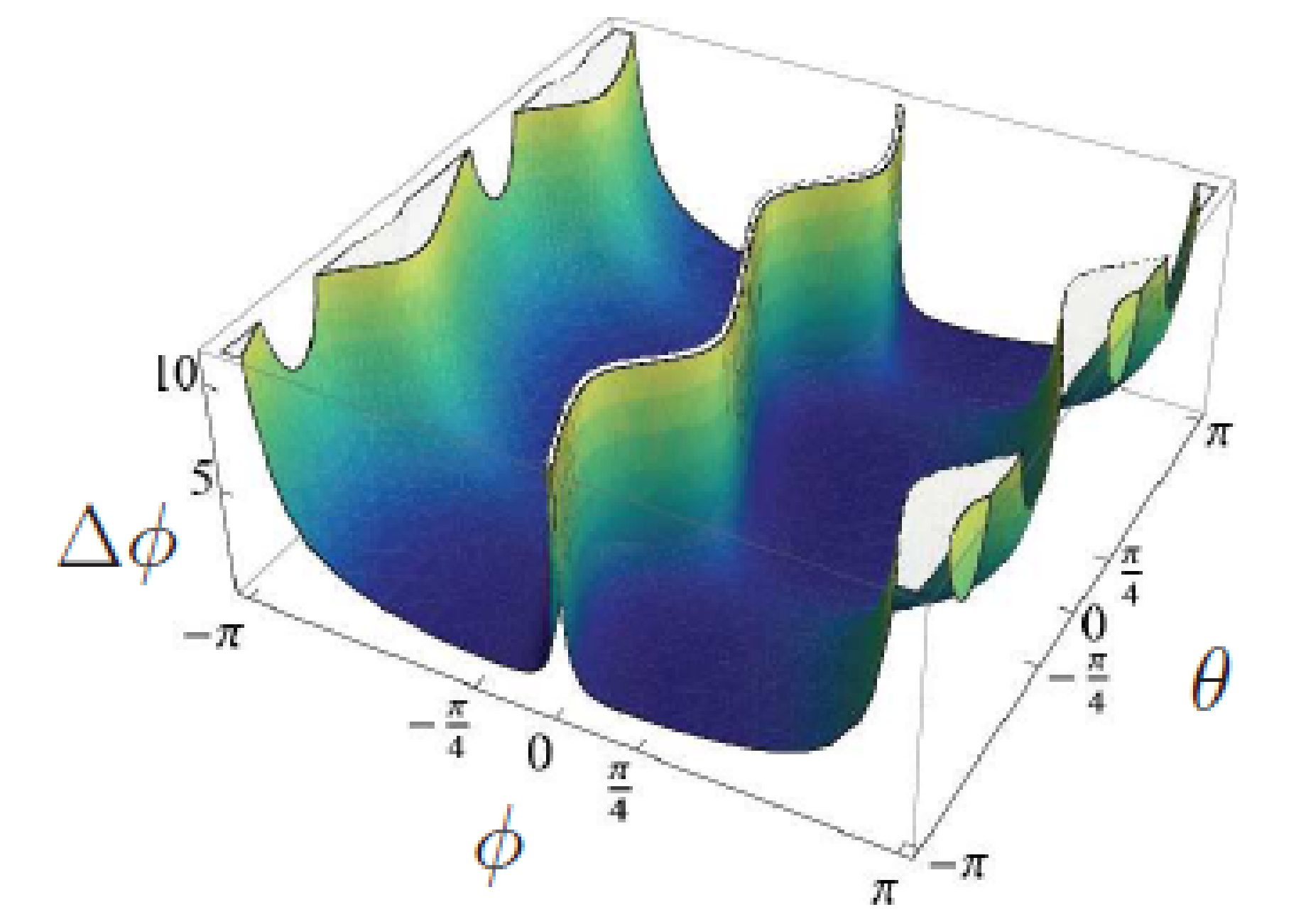}
\caption{\label{phase}A plot of the phase sensitivity as a function of the probe phase ($\phi$) and the input phase ($\theta$), with $r=0.5$ and the flux of the coherent input equal to what the squeezed flux would be with vacuum inputs.}
\end{figure}

\begin{figure}\centering
\includegraphics[scale=0.32]{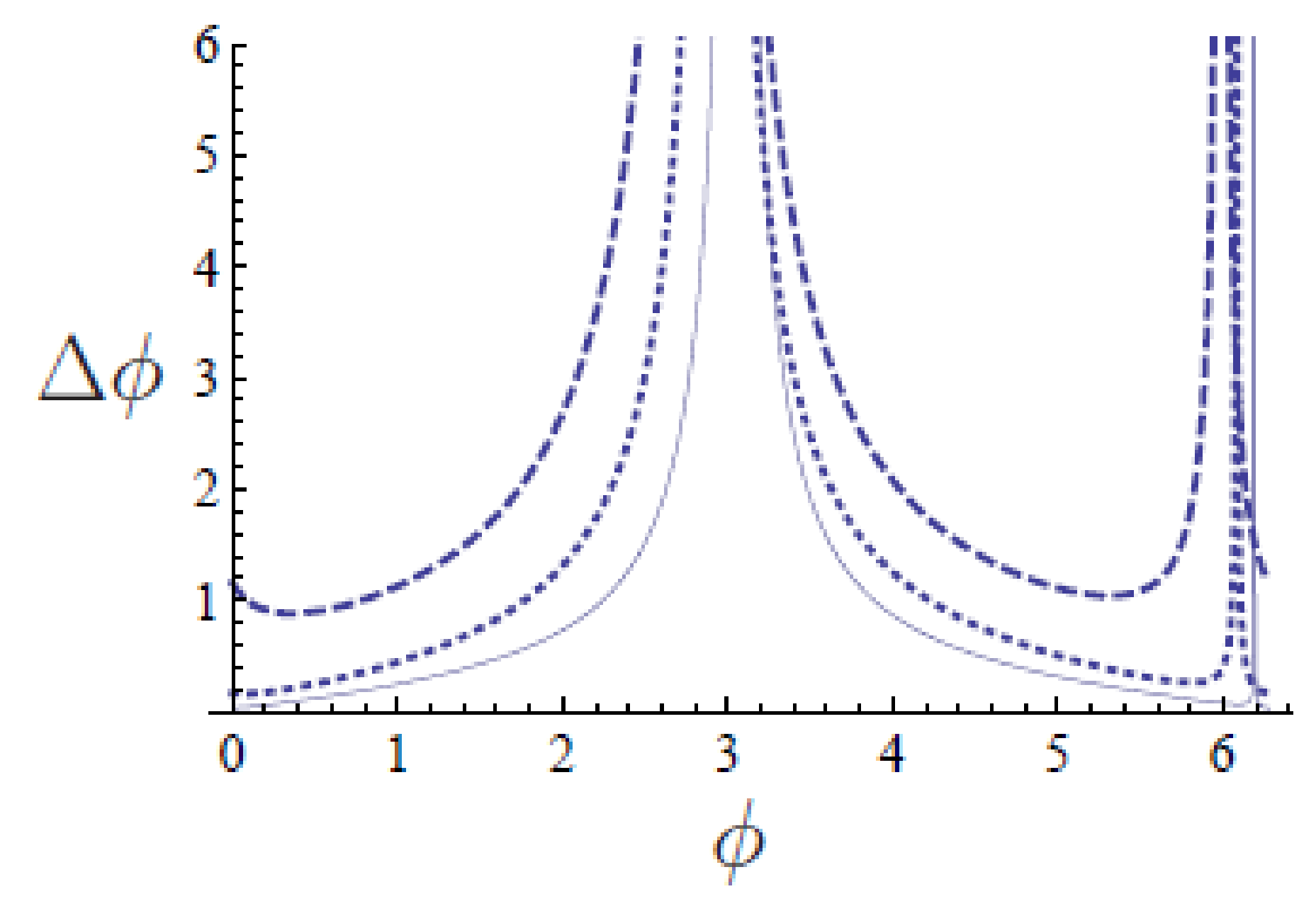}
\caption{\label{slices} Slices across Fig. \ref{phase} at input phase $\theta=\pi/4$. Phase sensitivity as a function of the probe phase, with the flux of the coherent input equal to what the squeezed flux would be with vacuum inputs. The gain ($r$) is set to 0.5 (dashed), 1 (dotted), and 1.5 (solid).}
\end{figure}

It should be noted that the result in Eq. (\ref{simple}), though simple, is not optimal for our scheme. The full expression Eq. (\ref{full}) shows that there is a complicated relationship between pump phase, probe phase, OPA gains, input coherent states, and the phase sensitivity. The choice of $\phi=0$, $\theta=\pi/4$ was made only because it caused significant simplification. If the amplitude of the input states and the gain of the OPAs are known, then the bias and input phases may be chosen such that the phase sensitivity is maximized. If this is done some additional minor improvement can be obtained. The effect is most pronounced at low photon numbers. Take as an example Fig. \ref{phase}; it is a plot of the phase sensitivity as a function of the probe phase ($\phi$) and the input phase ($\theta$), with $r=0.5$ and the flux of the coherent input equal to what the squeezed flux would be with vacuum inputs. For each value of $\theta$ there are values of $\phi$ for which the smallest detectable phase shift is minimized. These effects become less pronounced as either the gain or the coherent flux are increased. For clarity, slices at constant input phase ($\theta$) for various photon fluxes are provided in Fig. \ref{slices}. In \cite{klauder} it is shown that for an SU(1,1) interferometer with vacuum inputs the maximum phase sensitivity occurs at $\phi=0$, where all the light generated by the first four wave mixer is up-converted to the pump in the second (i.e. the output of the device as a whole is minimized). The relationship is more complicated in our case due to the interaction with the phase of the coherent inputs. However, by examining Figs. \ref{phase} and \ref{slices} we can see that the maximum phases sensitivity occurs generally near this value. On the other hand the setup is least sensitive to phase shifts near, and at $\phi=\pi$. This is where the output of the device as a whole is maximized.   

\section{Comparisons and Conclusions}

Let us take a simple numerical example. Gravitational wave detectors are essentially the largest and most sensitive interferometers to date. There are many of these interferometers around the world, the largest of which are in the LIGO project. Their interferometer arms have a circulating photon flux on the order of $10^{23}$photons/sec (20kW of circulating power at a wavelength of 1064nm \cite{power}). If we we take our theoretical setup with a gain of $r=3$ we can use Eq. (\ref{simple}) to calculate the coherent intensity necessary to achieve the same phase sensitivity (making the vast simplification that LIGO is a simple shot-noise-limited MZI) and arrive at $\sim2.5\times10^{18}$ photons/sec, forty thousand times less than LIGO. Vacuum inputs (squeezed light alone, as in the Klauder setup) would require a gain of $r\simeq 13.6$. Conversely if the coherent intensity were kept the same and the squeezed light added then the phase sensitivity would be improved by a factor of two hundred. Admittedly this is a large squeezing factor. However we would like to point out that a gain of $r=2.25$ has been reported recently, in \cite{bright} (a gain of 4.5 in the language of that paper). So it seems likely that a gain of three will be available in the near future. 

One could ask two pointed questions about our new scheme: (1) What advantage does this offer over techniques which are already planned for use in the next LIGO iteration and other next-generation, super-sensitive schemes; namely single-mode squeezed light incident into the dark input of an MZI? (2) In order to create bright entangled sources \--- such as the OPAs in our setup \--- one must employ a strong coherent pump beam. Is it really practically advantageous to employ this quantum light in an interferometer, rather than simply adding this strong pump laser into the device classically. In other words, taking into account the fact that very bright lasers are quite readily available, do squeezed light setups (ours in particular) really ``win out'' over the brute force technique of cranking up the coherent intensity as high as it will go?

\begin{figure}\centering
\includegraphics[scale=0.3]{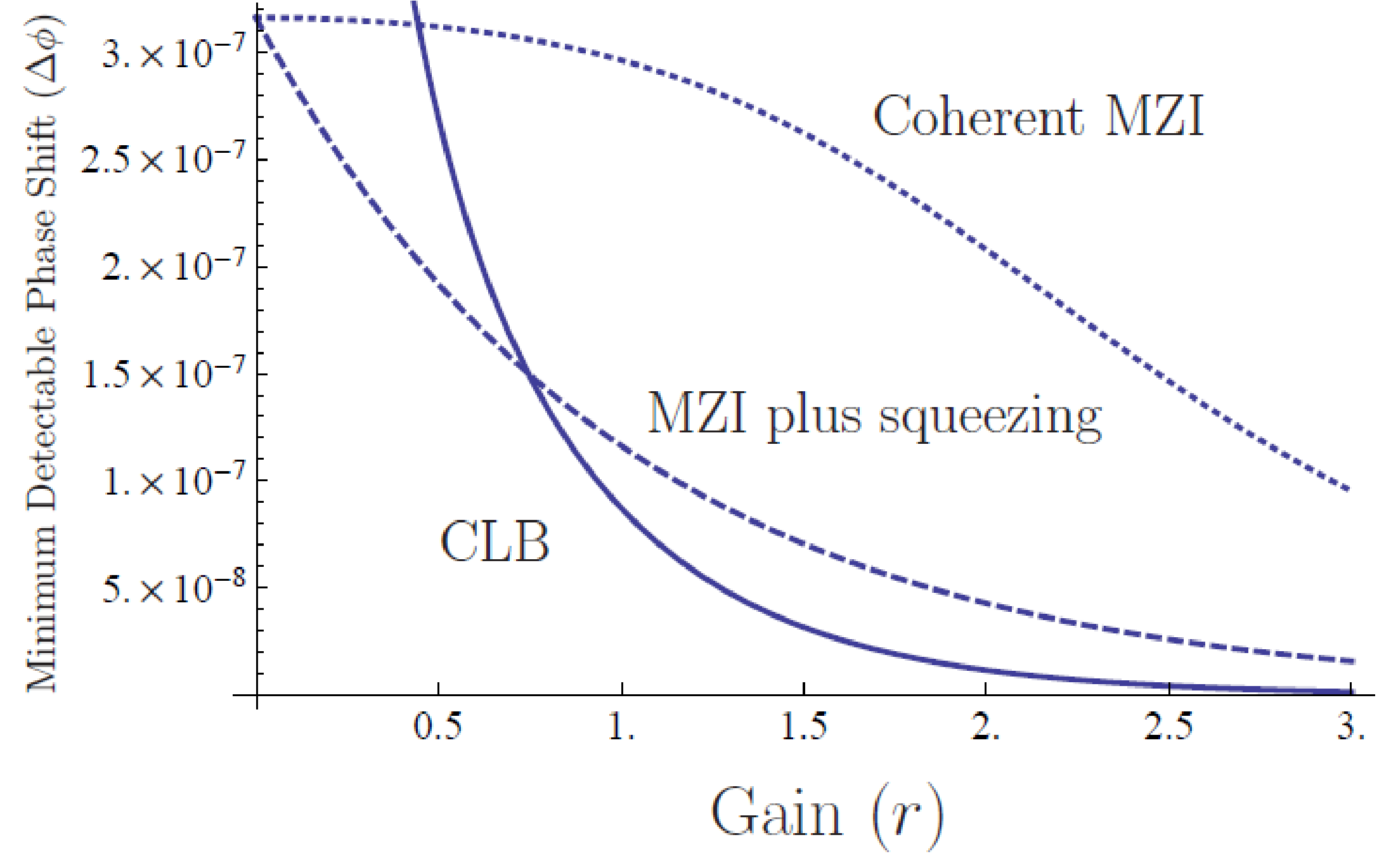}
\caption{\label{comp} Comparison of three interferometer schemes. The dashed line represents an MZI with $10^{13}$ photons of coherent light in one input mode and squeezed light in the other. The solid line is our coherent-light boost (CLB) scheme, with $5\times 10^{12}$ photons input into each mode. The dotted line is a standard coherent-input-only MZI with additional intensity proportional to what would be needed for a pump to create the equivalent gain in the other two setups: $I_{\mathrm{total}}=10^{13}+10^{12}\sinh^{2}(r)$. The critical point at which our scheme begins to do better than the single mode squeezed SU(2) (or coherent MZI) is given by $r=-\frac{1}{2}\mathrm{ln}2$ ($-\frac{1}{4}\mathrm{ln}2$ and is independent of the amount of coherent light (as long as it is equal between the two schemes). However, it should be pointed out that this graph uses the simple ``fixed phase'' version of our results in Eq. (\ref{simple}), which is most useful at higher gains. As discussed, at lower gains it becomes advantageous to optimize over the phases in Eq. (\ref{full}), gaining some small improvement in phases sensitivity beyond what is depicted in this figure. The asymptote at the origin remains, regardless \--- as in the limit of zero gain no mixing takes place between the modes. For values of $r$ below the critical point we may then say that the SU(1,1) does not do as well as the squeezed SU(2) (or coherent MZI) because the modes are only weakly mixed.} 
\end{figure}

The two questions are linked. To answer the second first: Super-sensitive interferometers suffer from both shot noise ($N_{S}$) and radiation-pressure noise ($N_{RP}$). At lower intensities $N_{S}$ dominates, however in the regime that LIGO now operates in, $N_{RP}$ has become as important, meaning that a kind of saddle point has been reached where either increasing or decreasing the intensity will lead to increased quantum noise (via $N_{RP}$ or $N_{S}$ respectively). Therefore it is very advantageous to use squeezed light as it reduces $N_{S}$ at a much faster rate than it increases $N_{RP}$. Furthermore, even just considering $N_{S}$, the rapid scaling of squeezed light does indeed ``win''. We can see this with a toy example. Suppose we have one interferometer which uses a coherent pump driven OPA to generate the squeezed light, and another which has a classical input with that same pump \textit{added} to the classical input beam (given that it takes approximately $10^{12}$ pump photons to make one pair of entangled photons \cite{1012}). This analysis is presented in Fig. (\ref{comp}). As can be seen, both a squeezed-light added MZI \--- and our coherent-light boosted setup \--- consistently outperform a solely coherent-input MZI for a wide range of gains.  

To answer the first question, take the high gain limit of Eq. (\ref{simple}) which reduces to $\Delta\phi=e^{-2r}/\sqrt{2N_{\mathrm{Coh}}}$. Compare this with a SU(2) MZI with one squeezed input (for $N_{\mathrm{Coh}}\gg N_{\mathrm{Squeezed}}$): $\Delta\phi=e^{-r}/\sqrt{N_{\mathrm{Coh}}}$ (see for example \cite{e2r}). Thus for large gains (a gain of $r=2$ is sufficient for being called large) our setup is exponentially better than single mode squeezing. Again we refer to Fig. (\ref{comp}) for a more complete comparison.

In conclusion we have presented a theoretical analysis of a new interferometric scheme that uses potentially strong coherent beams to boost the sensitivity of interferometers employing squeezed light. The result is a promising form of metrology, which achieves scaling far below the shot noise limit for bright sources. It uses only simple intensity measurement and components that are currently available. The phase sensitivity of this device scales as a coherent light input Mach-Zehnder interferometer multiplied by the sub-Heisenberg scaling of the Klauder et al. squeezed light interferometer. 

\section*{Acknowledgements}

W.N.P. would like to acknowledge the Louisiana Board of Regents and The Department of Energy for funding. Discussions with J. S. Kissel were invaluable. In addition J.P.D. acknowledges support from the Army Research Office, the Boeing Corporation, the Foundational Questions Institute, the Intelligence Advanced Research Projects Activity, and the Northrop-Grumman Corporation.


\section*{References}
\begin{thebibliography}{10}
\bibitem{Z}L. Zehnder, Z. Instrumentenkunde \textbf{11}, 275 (1891).
\bibitem{M}E. Mach, Z. Instrumentenkunde \textbf{12}, 89 (1892).
\bibitem{Budker}D. Budker and M. Romalis, Nature Phys. \textbf{3}, 227 (2007).
\bibitem{russ}M. E. Gertsenstein and V. I. Pustovoit, Sov. Phys. JETP \textbf{16}, 433 (1963). 
\bibitem{weiss}R. Weiss, Quarterly Progress Report, Research Laboratory of Electronics, MIT 105: 54 (1972). 
\bibitem{caves}C.M. Caves, Phys. Rev. D \textbf{23}, 1693 (1980).
\bibitem{kimble}Min Xiao, Ling-An Wu, H. J. Kimble, Phys. Rev. Lett. \textbf{59}, 278 (1987).
\bibitem{os}P. Grangier, R. E. Slusher, B. Yurke, and A. LaPorta, Phys. Rev. Lett. \textbf{59}, 2153 (1987).
\bibitem{luis}A. Luis, Phys. Lett. A \textbf{354}, 71 (2006).
\bibitem{kolkiran1}A. Kolkiran and G. S. Agarwal, Opt. Exp. \textbf{15}, 6798 (2007).
\bibitem{martini}F. Sciarrino, C. Vitelli, F. De Martini, R. Glasser, H. Cable, and J. P. Dowling, Phys. Rev. A \textbf{77}, 012324 (2008). 
\bibitem{hoffloss}Takafumi Ono, and H. F. Hofmann, (2009), arXiv:0910.3727v1 [quant-ph].
\bibitem{NdM}Chiara Vitelli, Nicol\`{o} Spagnolo, Lorenzo Toffoli, Fabio Sciarrino, and Francesco De Martini, (2010), arXiv:1004.2361v1.
\bibitem{boto}Agedi N. Boto, Pieter Kok, Daniel S. Abrams, Samuel L. Braunstein,
Colin P. Williams, and Jonathan P. Dowling, Phys. Rev. Lett. \textbf{85}, 2733 (2000).
\bibitem{agarwal}Girish S. Agarwal, Robert W. Boyd, Elna M. Nagasako, and Sean J. Bentley, Phys. Rev. Lett. \textbf{89}, 1389 (2001).
\bibitem{zei}Philip Walther, Jian-Wei Pan, Markus Aspelmeyer, Rupert Ursin,
Sara Gasparoni, and Anton Zeilinger, Nature \textbf{429}, 158 (2004).
\bibitem{steinberg}M. W. Mitchell, J. S. Lundeen, and A. M. Steinberg, Nature \textbf{429}, 161 (2004).
\bibitem{nagata}Tomohisa Nagata, et al., Science \textbf{316}, 726 (2007).
\bibitem{ono}Holger F. Hofmann, and Takafumi Ono, Phys. Rev. A \textbf{76}, 031806 (2007).
\bibitem{white}K. J. Resch, K. L. Pregnell, R. Prevedel, A. Gilchrist, G. J. Pryde, J. L. O'Brien, and A. G. White, Phys. Rev. Lett. \textbf{98}, 223601 (2007).
\bibitem{jon}Jonathan P. Dowling, Cont. Phys. \textbf{49}, 125 (2008).
\bibitem{gao}Yang Gao and Hwang Lee, J. Mod. Opt. \textbf{55}, 3319 (2008).
\bibitem{sean}Sean D. Huver, Christoph F. Wildfeuer, and Jonathan P. Dowling, Phys. Rev. A \textbf{78}, 063828 (2008).
\bibitem{ryan}Ryan T. Glasser, Hugo Cable, and Jonathan P. Dowling, Phys. Rev. A \textbf{78}, 012339 (2008).
\bibitem{vitelli}Chiara Vitelli, Nicolò Spagnolo, Fabio Sciarrino, and Francesco De Martini, J. Opt. Soc. Am. B \textbf{26}, 892 (2009).
\bibitem{Silb}I. Afek, O. Ambar, and Y. Silberberg, (2009), arXiv:0912.4009v2.
\bibitem{boixo2}Sergio Boixo, Animesh Datta, Steven T. Flammia, Anil Shaji, Emilio Bagan, and Carlton M. Caves, Phys. Rev. A \textbf{77}, 012317 (2008).
\bibitem{est}J. Est\`{e}ve, C. Gross, A. Weller1, S. Giovanazzi1, and M. K. Oberthaler, Nature \textbf{455}, 1216 (2008).
\bibitem{boixo1}Sergio Boixo, Animesh Datta, Matthew J. Davis, Anil Shaji, Alexandre B. Tacla, and Carlton M. Caves, Phys. Rev. A \textbf{80}, 032103 (2009).
\bibitem{higgins1}B. L. Higgins, D. W. Berry, S. D. Bartlett, M. W. Mitchell,
H. M. Wiseman, and G. J. Pryde, New J. Phys. \textbf{11}, 073023 (2009).
\bibitem{higgins2}B. L. Higgins, D. W. Berry, S. D. Bartlett, H. M. Wiseman, and G. J. Pryde, Nature \textbf{450}, 393 (2007).
\bibitem{kitten}Jonathan P. Dowling, Nature \textbf{450}, 362 (2007).
\bibitem{jeremy}Jeremy L. O'Brien, Science \textbf{318}, 1393 (2007). 
\bibitem{holland}M. J. Holland, and K. Burnett, Phys. Rev. Lett. \textbf{71}, 1355 (1993).
\bibitem{artur}Artur Widera, Olaf Mandel, Markus Greiner, Susanne Kreim,
Theodor W. H\"{a}nsch, and Immanuel Bloch, Phys. Rev. Lett. \textbf{92}, 160406 (2004).
\bibitem{qbs}Taesoo Kim, Jacob Dunningham, and Keith Burnett, Phys. Rev. Lett. \textbf{72}, 055801 (2005).
\bibitem{qbs2}J. Dunningham, and T. Kim, J. Mod. Opt. \textbf{53}, 557 (2006).
\bibitem{lee}Chaohong Lee, Phys. Rev. Lett. \textbf{97}, 150402 (2006).
\bibitem{kim}Taesoo Kim, and Heonoh Kim, J. Opt. Soc. Am. B \textbf{26}, 671 (2009).
\bibitem{p1}J. J. Bollinger, W. M. Itano, D. J. Wineland, and D. J.
Heinzen, Phys. Rev. A \textbf{54}, R4649 (1996).
\bibitem{p2}C. C. Gerry, Phys. Rev. A \textbf{61}, 043811 (2000).
\bibitem{p3}C. C. Gerry and R. A. Campos, Phys. Rev. A \textbf{64}, 063814 (2001).
\bibitem{p4}A. Chiruvelli and H. Lee, (2009), arXiv:0901.4395v1 [quant-ph].
\bibitem{p5}Y. Gao, C. F. Wildfeuer, P. M. Anisimov, H. Lee, and J. P. Dowling, (2009), arXiv:0907.2382v2 [quant-ph].
\bibitem{klauder}B. Yurke, S. L. McCall, and J. R. Klauder, Phys. Rev. A \textbf{33}, 4033 (1985).
\bibitem{eb}K. W\'{o}dkiewicz and J. H. Eberly, J. Opt. Soc. Am. B \textbf{2}, 458 (1985). 
\bibitem{Sanders}B. C. Sanders, G. J. Milburn, and Z. Zhang, J. Mod. Opt. \textbf{44}, 1309 (1997).
\bibitem{kolkiran2}A. Kolkiran and G. S. Agarwal, Opt. Exp. \textbf{16}, 6479 (2008).
\bibitem{in1}A. Zavatta, S. Viciani, and M. Bellini, Science \textbf{306}, 660 (2004).
\bibitem{in2}A. Zavatta, S. Viciani, and M. Bellini, Phys. Rev. A \textbf{72}, 023820 (2005).
\bibitem{in3}V. Parigi, A. Zavatta, Myungshik Kim, and M. Bellini, Science \textbf{317}, 1890 (2007).
\bibitem{NC}Available at http://math.ucsd.edu/~ncalg/.
\bibitem{power}B. Abbott, et al., Rep. Prog. Phys. \textbf{72}, 076901 (2009).
\bibitem{bright}I. N. Agafonov, M. V. Chekhova, and G. Leuchs, (2010), arXiv:0910.4831v3 [quant-ph].
\bibitem{1012}C. Kurtsiefer, M. Oberparleiter, and H. Weinfurter, Phys. Rev. A 64, 023802
(2001).
\bibitem{e2r}H. J. Kimble, Y. Levin, A. B. Matsko, K. S. Thorne, and S. P. Vyatchanin, Phys. Rev. D \textbf{65}, 022002 (2001).
\end{thebibliography}
\end{document}